\documentclass[12pt]{article}

\usepackage{epsf}
\usepackage[dvips]{graphicx}

\begin{document}

\begin{center}
{\bfseries Measurement of the 
tensor analyzing power 
$T_{20}$ in the   ${\rm dd\to {^3Hen}}$  and
${\rm dd\to {^3Hp}}$ at
intermediate energies
and at zero degree.
}

\vskip 5mm

\underline {V.P.~Ladygin$^{h,\dagger}$}, T.~Uesaka$^a$,   T.~Saito$^b$,
  M.~Hatano$^b$,
  A.Yu.~Isupov$^h$,
  H.~Kato$^b$,
  N.B.~Ladygina$^h$,
  Y.~Maeda$^a$,
  A.I.~Malakhov$^h$,
  J.~Nishikawa$^d$,
  T.~Ohnishi$^c$,
  H.~Okamura$^e$,
  S.G.~Reznikov$^h$,
  H.~Sakai$^{a,b}$,
  N.~Sakamoto$^c$,
  S.~Sakoda$^b$, 
  Y.~Satou$^f$,
  K.~Sekiguchi$^c$,
  K.~Suda$^a$,
  A.~Tamii$^f$,
  N.~Uchigashima$^b$ and
  K.~Yako$^b$
\vskip 5mm

{\small
$^a$ {\it
Center for Nuclear Study, University of Tokyo, Bunkyo, Tokyo 113-0033, Japan
}
\\
$^b$ {\it
Department of Physics, University of Tokyo, Bunkyo, Tokyo 113-0033, Japan
}
\\
$^c$ {\it
RIKEN, Wako, Saitama 351-0198, Japan
}
\\
$^d$ {\it
Department of Physics, Saitama University, Urawa 338-8570, Japan
}
\\
$^e$ {\it 
CYRIC, Tohoku University, Sendai, Miyagi 980-8578, Japan
}
\\
$^f$ {\it 
Research Center for Nuclear Physics, Osaka University, Ibaraki
567-0047, Japan
}\\
$^g$
{\it 
Department of Physics, Tokyo Institute of Technology, Tokyo 152-8551,
Japan
}\\
$^h$ {\it 
LHE-JINR, 141980, Dubna, Moscow region, Russia
}
\\

}
\end{center}

\vskip 5mm

\begin{center}
\begin{minipage}{150mm}
\centerline{\bf Abstract}
The data on the tensor 
analyzing power 
$T_{20}$ in the  $dd\to {\rm ^3He}n$ and $dd\to {\rm ^3H}p$  reactions 
 at 140, 200 and 270 MeV of the 
deuteron kinetic energy and at zero degree obtained
at RIKEN Accelerator Research Facility  
are presented. 
The observed positive  sign of $T_{20}$ 
clearly demonstrates the sensitivity 
to the D/S wave ratios in the ${\rm ^3He}$ and ${\rm ^3H}$ in the energy domain of the
measurements.
The $T_{20}$ data for the ${\rm ^3He}$-$n$ and ${\rm ^3H}$-$p$ channels are in
 agreement within experimental accuracy.

\vspace{1cm}

\vspace{0.5cm}

\end{minipage}
\end{center}

\vspace{4cm}

$^\dagger$ - Corresponding author:

Dr. V.P.~Ladygin,

LHE-JINR,

Joliot Curie 6,

141980, Dubna, Moscow region, Russia

phone: 7-09621-63929

FAX~ : 7-09621-65180

E-mail address: ladygin@sunhe.jinr.ru

\vskip 10mm

\newpage

\begin{sloppypar}

The spin structure of the light nuclei has been extensively investigated
 during the last decades using both electromagnetic and hadronic
 probes. 
The main purposes of these studies at intermediate and high energies are
to obtain the information on the high-momentum components of 
light nuclei affected by the relativistic effects and to search 
the manifestation of non-nucleonic degrees of freedom.
Especially, three nucleon bound states are of interest, because even a fundamental 
constant as the binding energy in the system 
cannot be reproduced 
by calculations with
modern pairwise nucleon-nucleon 
potentials~\cite{glock}.
Since the binding energy is known to have a strong correlation with strength 
of spin-dependent forces such as tensor force and/or three-nucleon force,
experimental study of the 
spin structure of three-nucleon bound system is expected to provide a clue to
understand the source of the missing energy.

The  non-relativistic Faddeev calculations~\cite{theory} for three-nucleon 
bound state predict
that the dominant components of the  ${\rm ^3He}$ ground state are a
symmetric $S$-state, where the ${\rm ^3He}$ spin due to the neutron  and 
two protons are in a spin singlet state and 
a $D$-state, where all three nucleon spins are oriented opposite to the 
${\rm ^3He}$ spin.
The $S$-state is found to dominate at small momenta
while $D$-state dominates at large momenta. 
The relative sign of $D$- and $S$- wave in the momentum space is positive 
at small and moderate nucleon momenta \cite{santos}.
\end{sloppypar}

\begin{sloppypar}

The sensitivity to the different components of ${\rm ^3He}$ can be observed in
polarization observables in both hadronic and electromagnetic
processes.

Polarized electron scattering on polarized ${\rm ^3He}$ target, 
${\rm \vec{^3He}(\vec{e},e')X}$,
can be used to study the different components of the 
${\rm ^3He}$ wave function \cite{theory}.
However, it is necessary to take into account final state interaction
 (FSI) and meson exchange currents (MEC) in addition to the  plane wave
 impulse approximation (PWIA) to describe  the experimental
 results~\cite{woodward} obtained at different
relative orientations of electron and ${\rm ^3He}$ spins.
Recent CEBAF data on the transverse asymmetry $A_{T^\prime}$~\cite{cebaf}
at $Q^2$ values of 0.1 and 0.2 GeV/c$^2$ have been 
described using full Faddeev calculation with the MEC effects.

The $\vec{\rm ^3He}(\vec{p},2p)$ and $\vec{\rm ^3He}(\vec{p},pn)$ 
breakup reactions were studied at TRIUMF
in quasielastic kinematics at  200 \cite{brash} and  290 MeV \cite{He3tr} of
incident proton energy.
In the last experiment, spin observables $A_{on}$, $A_{on}$ and $A_{nn}$ were 
measured up to
$q\sim 190$ and $\sim 80$ MeV/c for $\vec{\rm ^3He}(\vec{p},2p)$ and 
$\vec{\rm ^3He}(\vec{p},pn)$
reactions, respectively. The  results
indicate that analyzing powers $A_{no}$, $A_{on}$ and $A_{nn}$
are close to the PWIA   calculations for the 
${\rm \vec{^3He}(\vec{p},2p)}$ reaction,
while for  the  ${\rm \vec{^3He}(\vec{p},pn)}$
there is  a  strong disagreement with these predictions.
The same observables were recently measured at 197 MeV at IUCF 
Cooler Ring \cite{miller} up to
$q\sim 400$~MeV/c.
 It was observed that the polarization of the  neutron and  
proton at
zero nucleon momentum in ${\rm ^3He}$ are $P_n\sim 0.98$ and 
$P_p\sim -0.16$, respectively,
that is in good agreement
with the Faddeev calculations \cite{theory}.
However,
at higher momenta there is the discrepancy, which can be due to
the uncertainty of the theoretical calculations,
as well as to  large rescattering effects.

\end{sloppypar}

One Nucleon Exchange  (ONE) reactions, like ${\rm dp\to pd}$, ${\rm d{^3He}\to
p{^4He}}$ or
${\rm d^3He\to{^3Hed}}$, are the simplest processes with large momentum
transfer and, therefore, can be used as an effective tool to investigate the
structure of
the deuteron and ${\rm ^3He}$ at short distances.
In the framework of the ONE approximation \cite{vasan}  the polarization observables
of the above reactions are expressed in terms of the $D/S$- wave ratios of these
nuclei.
For instance, tensor analyzing power $T_{20}$ for
the  ${\rm dp\to pd}$ reaction in the collinear geometry is
expressed in terms of $D$ and $S$ wave of the  deuteron  in the momentum space
\begin{eqnarray}
\label{t20_zero}
T_{20} = \frac{ 2\sqrt{2}r -r^2}{\sqrt{2}(1+r^2)},
\end{eqnarray}
where $r$ is the $D/S$- wave ratio in the deuteron
at the corresponding internal momentum 
\cite{vasan}.

A  significant amount of the data devoted to the investigation
of the deuteron and ${\rm ^3He}$(${\rm ^3H}$) spin structure at short distances
have been accumulated in the last years.
Recently, the tensor analyzing power $T_{20}$ and polarization 
transfer coefficients
in backward elastic scattering, ${\rm dp\to pd}$, have been measured 
at Saclay, Dubna and RIKEN \cite{punjabi,azhg0,kimiko,sakai}.
Another binary reaction, ${\rm d^3He\to p^4He}$,  
has been investigated at RIKEN
using both polarized deuteron and ${\rm ^3He}$ up to 270 MeV \cite{tomo1,tomo2,tomo3}.
All the data show the sensitivity to the deuteron spin structure at short distances.
For instance, $T_{20}$ for both ${\rm dp\to pd}$ \cite{punjabi,kimiko,sakai} 
and ${\rm d{^3He}\to p{^4He}}$ \cite{tomo2,tomo3}
reactions at intermediate energies has a large negative value reflecting the
negative sign of the $D/S$- wave ratio in the deuteron in the momentum space.

As concerns ${\rm ^3He}$ spin structure,   
the tensor analyzing power $T_{20}$ in the 
$d{\rm ^3He}$ backward elastic scattering has been measured at 140, 200 and 270 MeV
\cite{tanifuji}. The sign of $T_{20}$ is found to be positive in accordance
with the positive sign of $D/S$- wave ratio in the ${\rm ^3He}$ \cite{santos}.

The data sensitive to the three-nucleon bound state spin structure are scarce,
and, new polarization data, especially, at short distances is of great importance.

The $dd\to {\rm ^3H}p({\rm ^3He}n)$ process is the simplest ONE reaction where three
nucleon structure is relevant.
 The theoretical analysis of the polarization phenomena for this reaction
in the collinear geometry
has been performed~\cite{lad1,nca_lad}. It has been 
shown that the tensor analyzing power $T_{20}$ due to
polarization of the incident deuteron can be expressed in the terms  
of the $D/S$- wave ratio in the ${\rm {^3H}({^3He})}$,
when three-nucleon bound state
is emitted in the forward direction in the cms.
A new experiment has been proposed \cite{r308n} to measure the energy and angular 
dependence of the tensor analyzing powers in the $dd\to {\rm ^3H}p({\rm
^3He}n)$ process at RIKEN.

In this paper the data on the tensor analyzing power $T_{20}$ due to
the incident deuteron polarization 
in the $dd\to {\rm ^3He(0^\circ)}n$ and $dd\to {\rm ^3H(0^\circ)}p$ reactions 
at 140, 200 and 270~MeV of the deuteron kinetic energy are presented.

\vspace{0.5cm}


The experiment has been performed at
RIKEN Accelerator Research Facility (RARF). 
A polarized deuteron
beam was produced by the high-intensity polarized ion source (PIS)
\cite{source} and accelerated by AVF 
and Ring cyclotrons up to the required energy.
The direction of the  beam polarization axis was controlled with a
Wien filter located at the exit of the PIS.
The polarization of the beam has been measured with 
the beam-line polarimeters based on the asymmetry measurements in ${\rm dp}$- elastic
scattering reaction, which has large values of tensor and vector analyzing powers
\cite{sak1,kimiko}.
New values of the analyzing powers for 
${\rm dp}$- elastic scattering at 140 and 270 MeV
obtained during absolute calibration of the
deuteron beam polarization via the ${\rm ^{12}C}(d,\alpha){\rm ^{10}B^*[2^+]}$
reaction 
\cite{suda} have been used.

SMART (Swinger and Magnetic Analyzer with a Rotator and a Twister)
spectrograph  \cite{smart} has been used for the measurements.
In this system, the incident beam can be swung by rotating the swinger magnet, 
while
the magnetic analyzer is fixed. 
The measurement of the particle's momentum and separation from the
primary beam was achieved by the magnetic system 
of SMART spectrograph consisting of two dipole and three quadrupole magnets
($Q$-$Q$-$D$-$Q$-$D$). 

The beam intensity during the
experiment was about 1-2 nA, the solid angle was $10^{-2}$ sr.
The live time of DAQ system \cite{acq} was
more than $80\%$ at the trigger rate of few thousands per second.
A CD$_2$ thin sheet \cite{maeda}
has been used as a deuterium target. 
The measurements on CD$_2$ and  carbon targets were made
for each setup setting to obtain the contribution from deuterium via
the CD$_2-$C subtraction. 
The thicknesses of the CD$_2$ and carbon targets used were
$54$~mg/cm$^2$ and $34$~mg/cm$^2$, respectively.

Three plastic scintillators viewed by photomultipliers 
from the both sides were used in coincidence for the trigger and to provide
the information about time-of-flight and energy losses of the particles.   
The distance between the target and detection point is about 17 m, which is
enough to separate tritons, deuterons and protons with the same momentum.
The time difference between the appearence of the trigger signal
and  radio-frequency signal of cyclotron was used as the time-of-flight information (TOF). 
The information about the charge of the particles was used at the trigger level.
Protons and deuterons were partly suppressed in the cases of ${\rm ^3He}$ 
and ${\rm ^3H}$ 
detection by  rasing threshold levels of the discriminators.

The event was used in the following  analysis only in the case when 
information on the pulse height in all tree
scintillation counters
was consistent with the energy losses for particle of interest.

The set of multiwire drift chambers has been used to obtain 
the information on the hit position in the focal plane and, therefore,
on the momentum of the particle and emission angle from the target.
The energy resolution provided by the tracking system was 
$\sim 300$~keV. The typical track reconstruction efficiency   was better
than 99\%.

In the experiment the  data for
${\rm ^3He}$-$n$ channels have been 
obtained at 140, 200 and 270 MeV, while for 
the ${\rm ^3H}$-$p$
channel at 140 and 200~MeV only, because the momentum of the ${\rm ^3H}$
at 270 MeV is higher than the maximal rigidity of SMART \cite{smart}.

The quality of the  CD$_2-$C subtraction procedure 
for the $dd\to{\rm ^3He(0^\circ)}n$ 
reaction  at 270, 200 and 140~MeV is demonstrated in Fig.1 a), b) and c),
respectively.
The spectra are plotted versus
excitation energy $E_X$ defined as following
\begin{eqnarray}
E_X=\sqrt{((E_0-E_{3N})^2-({\bf P}_0-{\bf P}_{3N})^2}-M_{N},
\end{eqnarray}
where {\bf P$_0$} is the incident momentum; $E_0=2M_d+T_d$ is the total initial energy;  
$E_{3N}$ and {\bf P$_{3N}$}  are energy and  momentum  of the 
three-nucleon system, respectively; $M_{N}$ is the nucleon mass.
Peaks at $E_{X}=0$~MeV
correspond to ${\rm ^3He}$ from the $dd\to {\rm ^3He}n$ reaction,
while the shadowed areas show the carbon contribution.

For the ${\rm ^3H}$ detection case 
the yield from carbon under peak at $E_X=0$~MeV is negligibly
small.
The peak for the binary reaction on deuterium, $dd\to{\rm ^3H}p$, is separated
from the peaks for the reaction $d^{12}C\to{\rm ^3H}X$ by $\sim 5$ and $\sim 10$~MeV
at 200 and 140~MeV, respectively. 

The events considered for the analysis were selected 
within polar angle acceptance $\le 1.4^\circ$.

Only the events for 
unpolarized spin mode and two spin modes with the 
ideal values
of the tensor polarization $p_{zz}=-2$ and $p_{zz}=+1$ \cite{source} 
were used for the analysis. Actual  
values of the beam polarization were $\sim$75\%, $\sim$50\%  and $\sim$15\% of the
ideal values at 270~MeV, 200~MeV and 140~MeV, respectively.

Tensor analyzing power $T_{20}$ for each polarized spin mode
 was calculated  from the following expression
\begin{eqnarray}
T_{20}= -\frac{2\sqrt{2}}{p_{ii}} \left ( \frac{\sigma_{pol}}{\sigma_0}-1 \right ),
\end{eqnarray}
where $p_{ii}$ is the corresponding tensor polarization of the beam,
$\sigma_{pol}$ and $\sigma_0$ are yields 
in polarized and unpolarized spin modes obtained via 
CD$_2-$C subtraction after  corrections of  dead-time, 
detection efficiency and beam intensity.
The analyzing power $T_{20}$ was taken as weighted averaged for two
spin modes.

\vspace{0.5cm}

The results on the tensor analyzing power  $T_{20}$ for the
$dd\to {\rm ^3He(0^\circ)}n$ and $dd\to {\rm ^3H(0^\circ)}p$ reactions
are given in the Table 1. 
The systematic error due to uncertainty in  the beam
polarization and statistical error are added in quadrature.

These data are also shown in Fig.2.
The open triangles and full circles correspond to the ${\rm ^3H}$-$p$ and  
${\rm ^3He}$-$n$ channels, respectively. 
The $T_{20}$ values obtained for the both charge symmetrical 
${\rm ^3H}$-$p$ and ${\rm ^3He}$-$n$ channels at 140 and 200~MeV in this
experiment are in good agreement within achieved experimental accuracy.
No evidence of possible charge symmetry breaking 
is found in these processes.

The positive sign of $T_{20}$ values is in a striking contrast to  negative $T_{20}$
for 
${\rm dp\to pd}$ or other reactions where deuteron structure is relevant.
The  energy dependence of $T_{20}$ demonstrates
the increasing of $T_{20}$ magnitude with the energy. 
This behaviour can be understood in terms of $D/S$ wave ratio in the
${\rm ^3He(^3H)}$ with the help of ONE.  

Within ONE approximation tensor analyzing power $T_{20}$ for
the  $dd\to{\rm ^3H}p({\rm ^3He}n)$ reaction in the collinear geometry is
expressed in terms of $D$ and $S$ wave ratio $r$ of the  ${\rm ^3H(^3He)}$ 
\cite{lad1,nca_lad,r308n} (see expression (\ref{t20_zero})).
The positive sign of  $T_{20}$ in the explored energy domain reflects
the  positive sign of the $D/S$ wave ratio in the ${\rm ^3He}$(${\rm
^3H}$) in the momentum space \cite{theory,santos}. 
In this respect one can conclude that our data are sensitive to
the $D$-state in the ${\rm ^3He(^3H)}$.

The solid, dashed and dotted curves in Fig.~2 
are the results of non-relativistic  ONE calculations
\cite{r308n} using Urbana \cite{urbana}, Paris \cite{laget5} and
RSC \cite{rsc} (with the parametrization from \cite{uzikov}) ${\rm ^3He}$ wave
functions.
Paris deuteron wave function \cite{paris} was used in the calculations.  
The data are in a qualitative agreement with the ONE calculations
\cite{r308n}.  

The behaviour of our data is consistent with the behaviour of $T_{20}$ for
other reactions sensitive to the ${\rm ^3He}$
spin structure. In Fig.3 the 
data on $T_{20}$ in the  $dd\to {\rm ^3He(0^\circ)}n$ are plotted 
versus internal momentum $k$ along with
the data obtained for the $d{\rm ^3He}\to {\rm ^3He}d$ reaction \cite{tanifuji}.
The data for the both processes 
demonstrate the universality in the $k$- behaviour. 
The solid curve is the result of ONE calculation
using Urbana ${\rm ^3He}$ wave function \cite{urbana} according to Eq.(\ref{t20_zero}).
The relativistic effects are taken into account by the minimal relativization scheme 
\cite{relat}.

 The discrepancy between the data and the calculations can be due 
 both to the contribution from the reaction mechanism other 
 than ONE and to the nonadequate description of the short-range 
 ${\rm ^3He}$ spin structure. Concerning the reaction mechanism, the virtual
excitation to the other channels, for example, excitation to 
$\Delta$-isobar, is considered phenomenologically in Ref.~\cite{tanifuji}
to reproduce the energy dependence of $T_{20}$ for the $d{\rm ^3He}\to {\rm ^3He}d$
process.
The microscopic calculation by Laget et al.\cite{laget5} shows that the
 $\Delta$-isobar contribution to the $dd\to{\rm ^3He}n$ process is less than 10\% at
energies lower than 300~MeV.
At higher energies, in GeV region, it is shown that the coherent
sum of ONE and the $\Delta$-isobar excitation reproduces the cross section
data reasonably. Thus it is expected that the uncertainty in the
reaction mechanism is too small to explain the descrepancy between
the data and the ONE calculation. Further theoretical investigations
of the short-range ${\rm ^3He}$ spin structure as well as the reaction 
mechanism of the $dd\to {\rm ^3He}n$ process are needed to understand the
$T_{20}$ data presented here.

The extension of the $T_{20}$ measurements to the higher energies, 
namely to larger internal momenta, is of great interest.
In particular, the measurement of $T_{20}$ in the vicinity of $k\sim$0.4~GeV/c,
where the changing of the $T_{20}$ sign is expected, could distinguish 
different models of the short range ${\rm ^3He}$ spin structure description.

\vspace{0.5cm}

The results can be summarized as following.

The data on the tensor analyzing power 
$T_{20}$ in the $dd\to {\rm ^3H}p$ and $dd\to {\rm ^3He}n$ 
reactions at intermediate energies and 
in collinear geometry are obtained.
The sign of $T_{20}$ is positive being in the agreement with the
results on $T_{20}$ in the $d{\rm ^3He}\to {\rm ^3He}d$ reaction \cite{tanifuji}, on
the
one hand, and with the ONE calculations using standard ${\rm ^3He}$ wave functions,
on the other hand. 

In general, ONE reproduces the global feature of the $T_{20}$ 
energy dependence. 
However, the deviation of the experimental data from the
ONE calculations can be due to not only the nonadequate
description of the short range ${\rm ^3He}$ 
spin structure, but also the influence of the mechanisms additional to ONE. 
To improve the description  of the obtained data further theoretical
calculations are required.

The extension of $T_{20}$ measurements  to a GeV energy range 
is important  to explore the short range ${\rm ^3He}$ spin structure.

\vskip 5mm
The authors express their thanks to the staff of RARF for
providing of excellent conditions for the R308n experiment.
They are grateful to  H.Kumasaka, R.Suzuki and R.Taki for the help
during experiment.
Russian part of collaboration thanks RIKEN Directorate for kind
hospitality during the experiment.
The work has been supported in part by the JINR-Bulgaria  grants for 2001 and 2002
years and by the Russian 
Foundation for Fundamental Research (grant $N^o$ 04-02-17107).

\newpage

~
\vspace{2cm}

Table 1. 
Tensor analyzing power $T_{20}$ in the ${\rm dd\to{^3Hen}}$ and ${\rm dd\to{^3Hp}}$
reactions.

\begin{center}
\begin{tabular}{|c|c|c|c|}
\hline
Energy, MeV & Reaction &  $T_{20}$ & $\Delta T_{20}$ \\
\hline
140 & ${\rm dd\to{^3Hen}}$& 0.112 & 0.019  \\
140 & ${\rm dd\to{^3Hp}}$ & 0.082 &  0.018 \\
\hline
200 & ${\rm dd\to{^3Hen}}$ & 0.172 & 0.020 \\
200 & ${\rm dd\to{^3Hp}}$  & 0.165 & 0.018 \\
\hline
270 & ${\rm dd\to{^3Hen}}$ & 0.300 & 0.013 \\
\hline
\end{tabular}
 
\end{center}

\newpage

~
\vspace{1cm}

\centerline{\bf Figures caption}

\vspace{1cm}

{\bf Fig.1.} 
$CD_2-C$ subtraction for the $dd\to^3He(0^\circ)n$ channel:
a) at 270 MeV, b) at 200  MeV and c) at 140 MeV. The open and shadowed histograms
correspond to the yields from $CD_2$ and carbon targets, respectively.
\vspace{0.5cm}

{\bf Fig.2.} 
Tensor analyzing power $T_{20}$ 
in the  ${\rm dd\to{^3Hen}}$(solid circles) and  ${\rm dd\to {^3Hp}}$ (open triangles)
reactions in collinear geometry versus incident momentum of deuteron. 
The solid, dashed and dotted curves are the results of the non-relativistic ONE
calculations
\cite{r308n} for the forward emission of the ${\rm {^3He}({^3H})}$ 
in the cms, using Urbana \cite{urbana}, Paris \cite{laget5} and RSC \cite{rsc} 
 ${\rm ^3He}$ wave functions, respectively. Paris deuteron wave function
\cite{paris} was used
for the deuteron structure description.
\vspace{0.5cm}

{\bf Fig.3.} 
Tensor analyzing power $T_{20}$ 
in the  ${\rm dd\to{^3Hen}}$(solid circles) and  ${\rm d{^3He}\to {^3He}d}$ (open
squares)
\cite{tanifuji} 
reactions versus internal momentum $k$. 
The solid curve is the result of the ONE calculations
using Urbana  ${\rm ^3He}$
wave function \cite{urbana}. 

\newpage

~
\vspace{1cm}

\begin{figure}[hbt]
 \epsfxsize=120mm
 \epsfysize=120mm
 \centerline{
 \epsfbox{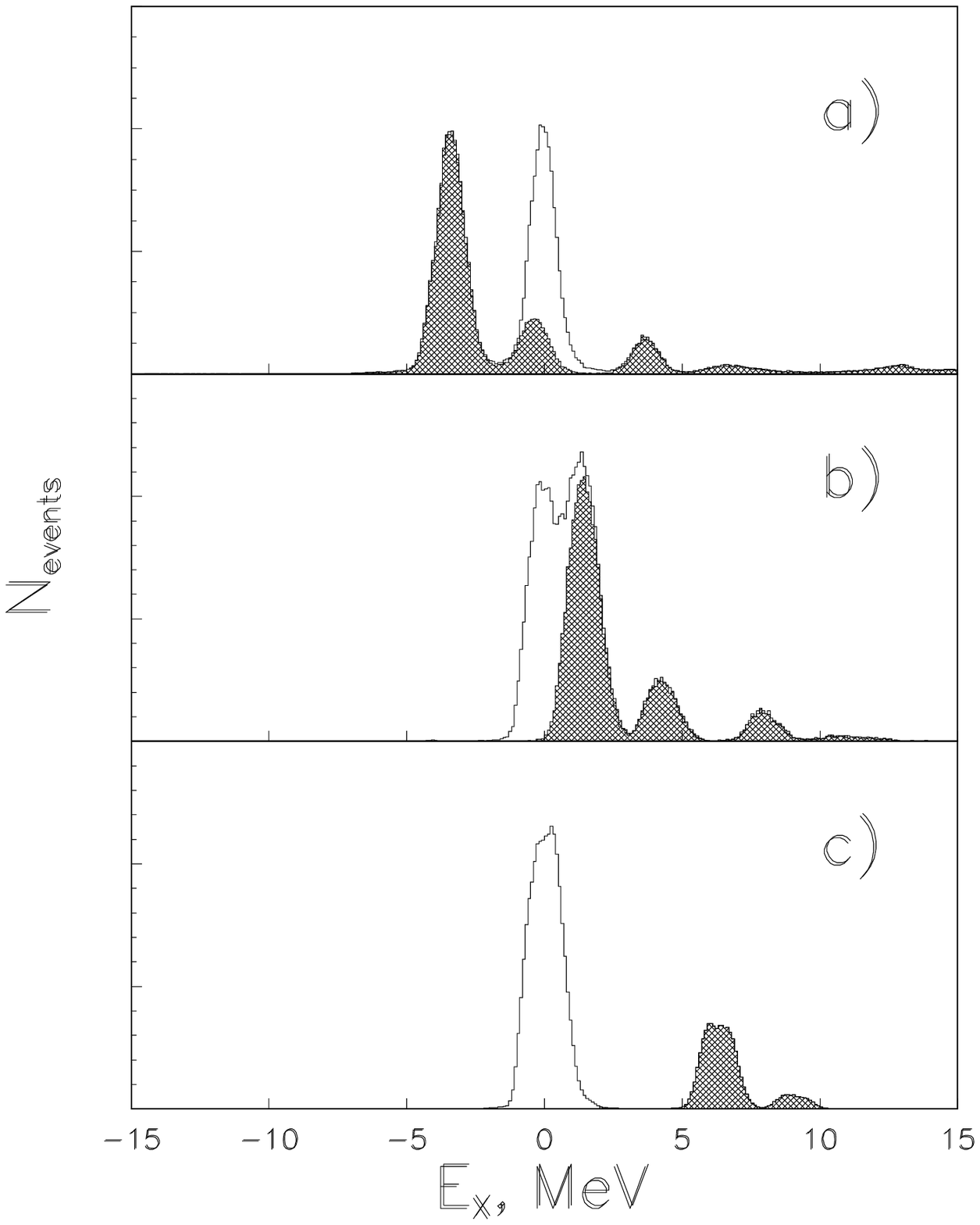}}
\caption{~~~~}
\end{figure}

\newpage

~
\vspace{2cm}

\begin{figure}[hbt]
 \epsfxsize=100mm
 \epsfysize=100mm
 \centerline{
 \epsfbox{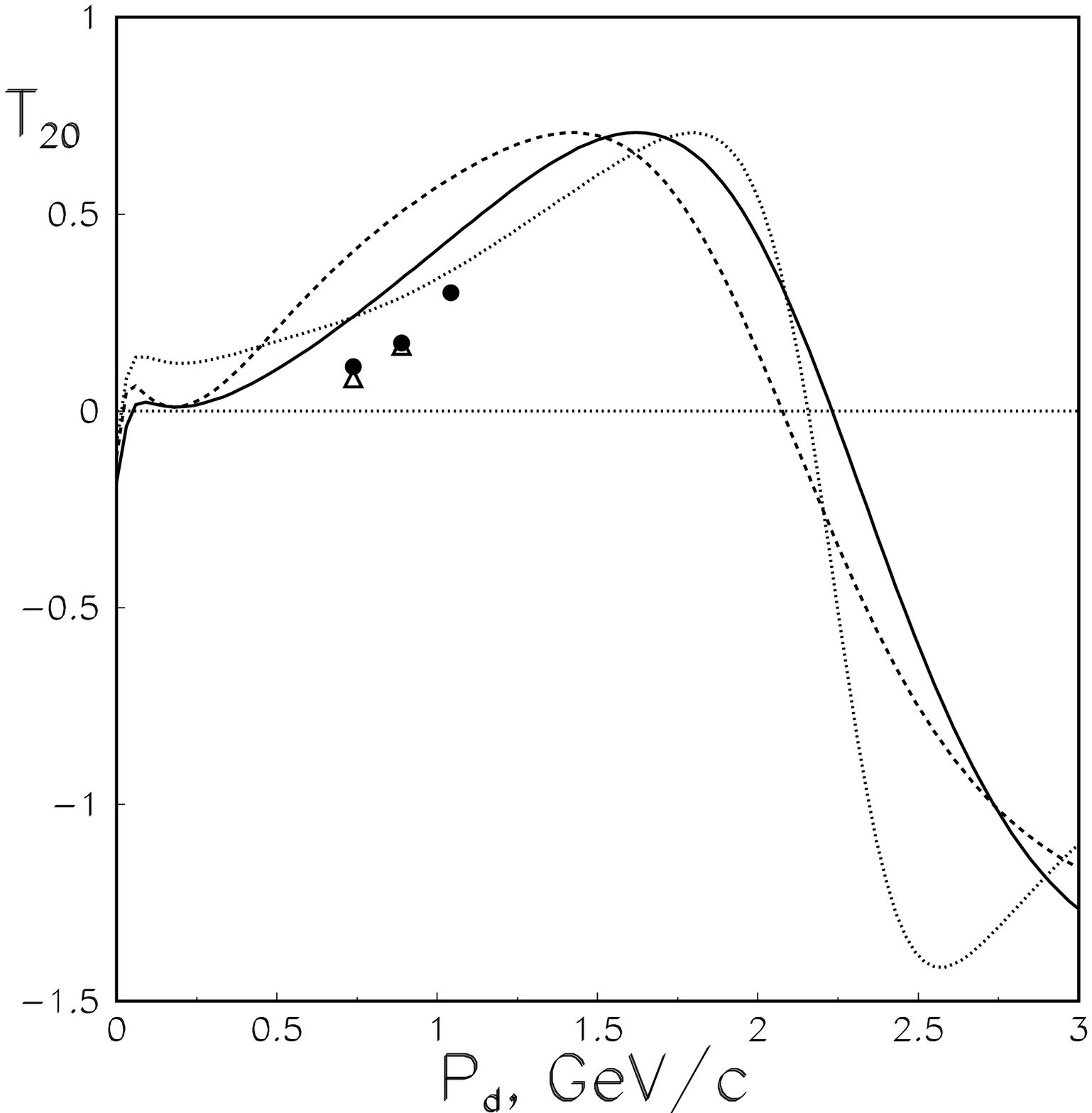}}
\caption{~~~~}
\end{figure}

\newpage

~
\vspace{2cm}

\begin{figure}[hbt]
 \epsfxsize=100mm
 \epsfysize=100mm
 \centerline{
 \epsfbox{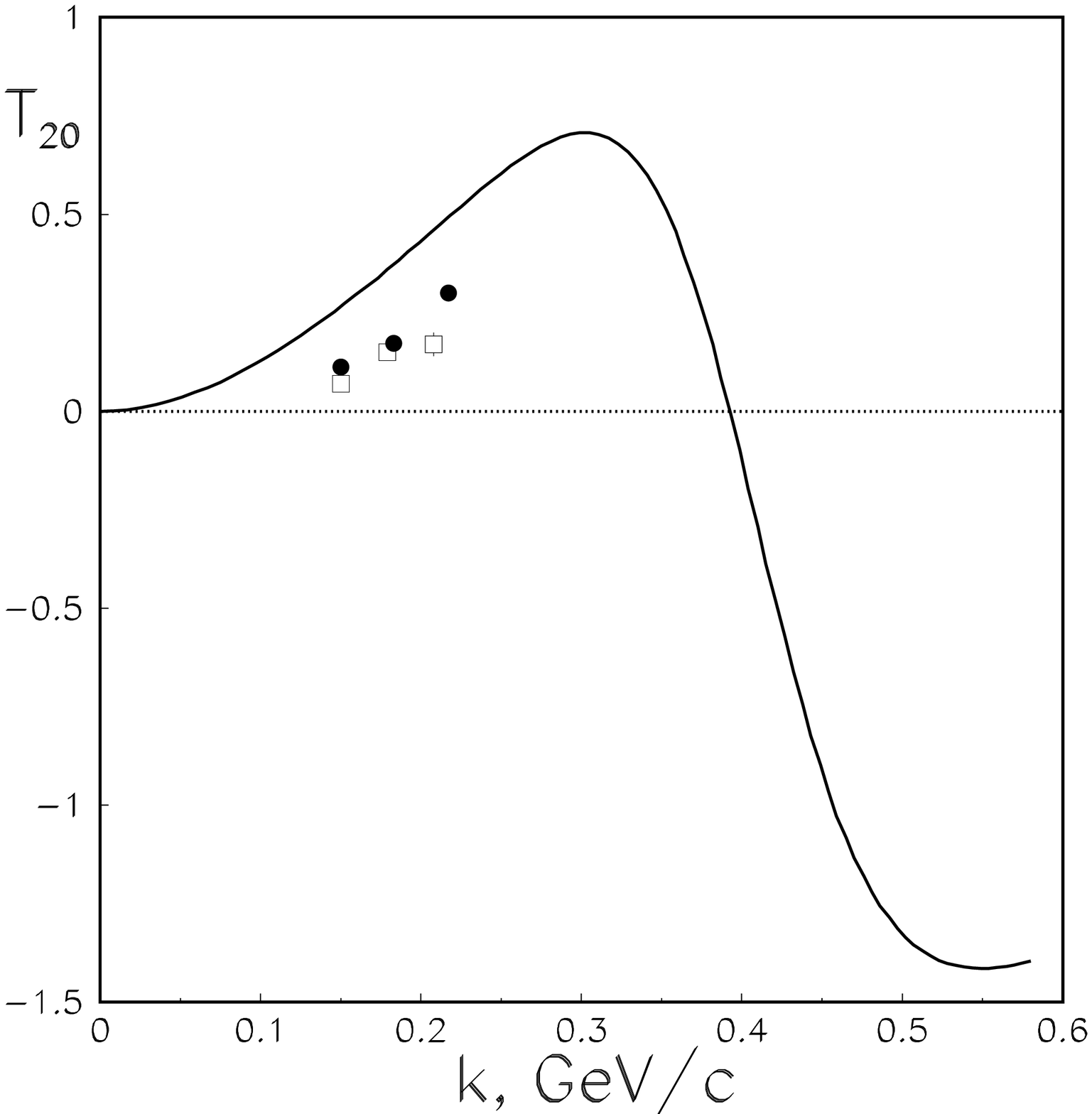}}
\caption{~~~~ }
\end{figure}

\end{document}